\documentclass[Journal]{IEEEtran}
\ifCLASSINFOpdf

\else

\fi

\usepackage{cite}
\usepackage{amsmath,amssymb,amsfonts}
\usepackage{algorithmic}
\usepackage{graphicx}
\usepackage{textcomp}
\usepackage{xcolor}
\usepackage{mathrsfs}
\usepackage{amsmath}
\usepackage{bm}
\usepackage{graphicx}
\usepackage{float} 
\usepackage{subfigure}



\begin{document}
\begin{sloppypar}

\title{Resource Allocation for Text Semantic Communications}

\vspace{-15pt}
\author{\IEEEauthorblockN{Lei Yan, Zhijin Qin,~\IEEEmembership{Senior Member,~IEEE}, Rui Zhang,~\IEEEmembership{Member,~IEEE}, Yongzhao Li,~\IEEEmembership{Senior Member,~IEEE}, and Geoffrey Ye Li,~\IEEEmembership{Fellow,~IEEE}}
\vspace{-20pt}
\thanks{Lei Yan, Rui Zhang, and Yongzhao Li are with
the State Key Laboratory of Integrated Services Networks, Xidian University, Xi’an 710071, China (e-mail: lyan@stu.xidian.edu.cn; rz@xidian.edu.cn; yzhli@xidian.edu.cn).

Zhijin Qin is with the School of Electronic Engineering and Computer Science, Queen Mary University of London, London E1 4NS, U.K. (e-mail: z.qin@qmul.ac.uk). Geoffrey Ye Li is with School of Electrical and Electronic Engineering, Imperial College London, London SW7 2AZ, UK (e-mail: geoffrey.li@imperial.ac.uk).
}
}
\maketitle

\vspace{-30pt}
\begin{abstract}
Semantic communications have shown its great potential to improve the transmission reliability, especially in the low signal-to-noise regime. However, resource allocation for semantic communications still remains unexplored, which is a critical issue in guaranteeing the semantic transmission reliability and the communication efficiency. To fill this gap, we investigate the spectral efficiency in the semantic domain and rethink the semantic-aware resource allocation issue. Specifically, taking text semantic communication as an example, the semantic spectral efficiency (S-SE) is defined for the first time, and is used to optimize resource allocation in terms of channel assignment and the number of transmitted semantic symbols. Additionally, for fair comparison of semantic and conventional communication systems, a transform method is developed to convert the conventional bit-based spectral efficiency to the S-SE. Simulation results demonstrate the validity and feasibility of the proposed resource allocation method, as well as the superiority of semantic communications in terms of the S-SE.
\end{abstract}
\vspace{-5pt}

\vspace{-15pt}
\section{Introduction}
\vspace{-3pt}
With growing wireless applications and increasing data traffic, wireless communications are facing the bottleneck of spectrum scarcity, which motivates a paradigm shift from conventional to semantic communications\cite{tong2021challenges,zhijin2021challenges}. By focusing on transmitting the meaning of the source, semantic communications have shown a great potential to reduce the network traffic and thus alleviate spectrum shortage. Particularly, different types of semantic systems have been studied for different types of sources, including text\cite{DeepSC,sana2021learning}, image\cite{SemanticImage,JSCC-image}, speech\cite{DeepSC-S}, and video\cite{tung2021deepwive}, to ensure significant improvement in semantic transmission reliability. In this context, it is vital to investigate the resource allocation issue for semantic communications to improve the communication efficiency while guaranteeing the transmission reliability \cite{semantic-empowered}.

In wireless communications, how to measure the information content as well as the spectral efficiency (SE) is fundamental to the resource allocation issue. Bit is used in the conventional communications. However, it is not applicable in semantic communications as bits are produced based on the statistic knowledge of source symbols rather than the semantic information of the source. Therefore, resource allocation needs to be rethought from the semantic perspective. The research on semantic theory has provided some insights on this issue. Carnap \cite{logic} first attempted to measure the semantic information in a sentence based on the logical probability. On this basis, the semantic channel capacity was derived in \cite{semanticLimitation} for the discrete memoryless channel, revealing the existence of the semantic coding strategy for reliable communications. Furthermore, semantic coding, the fundamental limits of semantic transmission, and semantic compression were investigated in \cite{semanticCompression}. However, the aforementioned works are based on abstract models without any hint of practical implementation and fail to quantify the SE in the semantic domain.

Although a complete theory or a well-developed mathematical model for semantic communications is still missing, the success of semantic system design with the aid of deep learning (DL) makes it possible to define a calculable SE in the semantic domain. Particularly, the DL-enabled semantic communication system (DeepSC) \cite{DeepSC} and its several variants\cite{sana2021learning,VQA} can effectively extract the semantic information from text and successfully deliver the meaning to the receiver.
In this article, we use DeepSC as an example to explore the SE issue and the resource allocation problem in such a semantic-aware network. The main contributions are as follows:
\begin{itemize}
    \item A novel resource allocation model is proposed for semantic-aware networks. Specifically, the semantic spectral efficiency (S-SE) is first defined to measure the communication efficiency from the semantic perspective. Then a new formulation is proposed and solved to maximize the overall S-SE in terms of channel assignment and the number of transmitted semantic symbols. 
    \item To make a fair comparison between semantic and conventional communication systems, a transform method is developed to convert the bit-based SE to the S-SE.
    \item Simulation results verify the effectiveness of the proposed resource allocation model, as well as the superiority of semantic communication systems in terms of the S-SE.
\end{itemize}

The rest of this article is organized as follows. Section II introduces the system model. Semantic-aware resource allocation is formulated and solved in Section III. Section IV introduces a transform method for fair comparison of semantic and conventional communication systems and presents the simulation results. Section V concludes the article.

\textit{Notation:} $\mathbb{R}^{n \times m}$ represents the set of real matrices of size $n \times m$. Bold-font variables represent matrices and vectors. $x\sim{\mathcal{CN}(\mu,\sigma^2)}$ means $x$ follows a circularly-symmetric complex Gaussian distribution with mean $\mu$ and covariance $\sigma^2$.

\vspace{-10pt}
\section{System Model}
\vspace{-3pt}
We consider a cellular network consisting of a base station (BS) and a set of users denoted by $\mathcal{N}=~\{1,2,\dots,n,\dots,N\}$, as shown in Fig.~1. %Semantic communication technique is utilized in the uplink transmission.
%to improve transmission reliability and communication efficiency. Particularly, 
DeepSC\cite{DeepSC} is adopted as the semantic communication model and equipped at each user for text transmission, where the semantics underlying text can be effectively extracted through Transformer. The DeepSC transceiver is assumed to be trained at the BS or cloud platforms. Then the trained semantic transmitter model is broadcast to users. In the following, we will detail the DeepSC transmitter at users, the transmission model, and the DeepSC receiver at the BS.
\begin{figure}
\vspace{-4pt}
  \centering
  \includegraphics[width=0.30\textwidth]{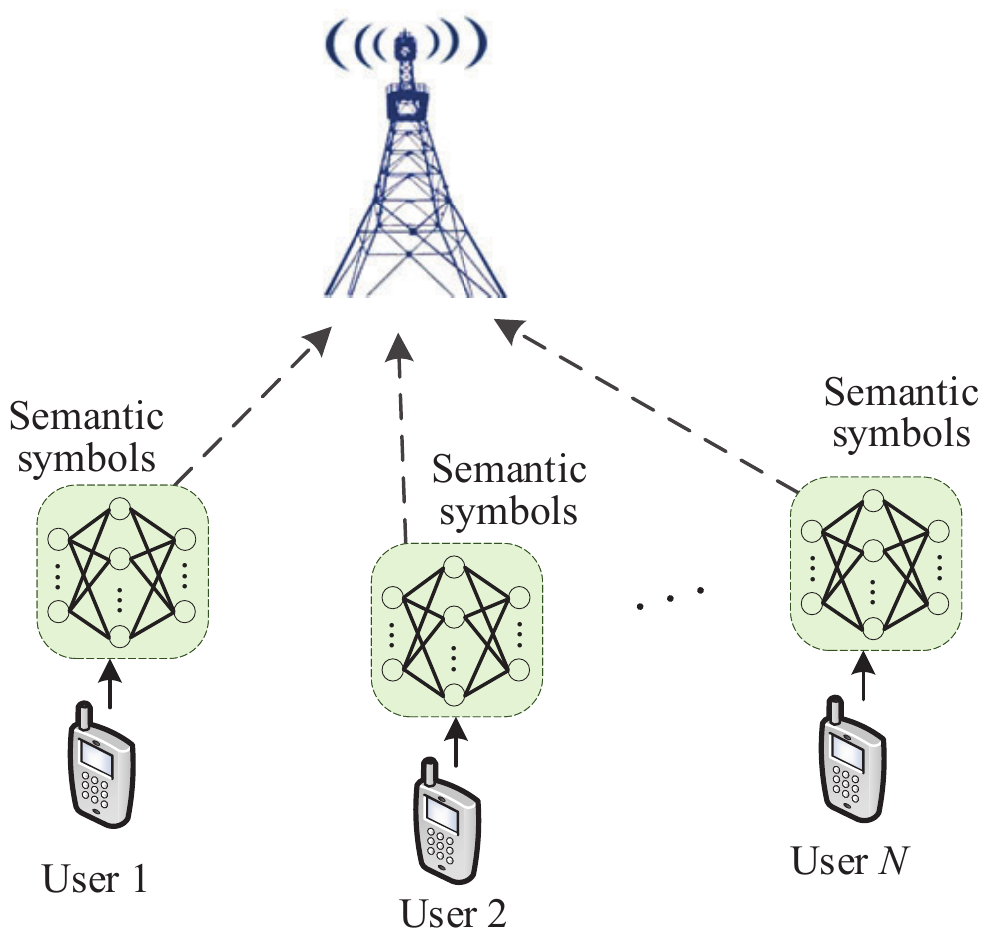}
\vspace{-10pt}
  \caption{The structure of semantic-aware networks.}
 \vspace{-18pt}
\end{figure}
\vspace{-10pt}
\subsection{DeepSC Transmitter}
\vspace{-2pt}
In our model, the $n$-th user generates a sentence  ${\bf{s}}_n=[w_{n,1}, w_{n,2}, \dots,w_{n,l}, \dots, w_{n,L_n}]$, where $w_{n,l}$ denotes the $l$-th word and $L_n$ is the sentence length at the $n$-th user. Then the sentence is fed into the DeepSC transmitter and mapped to a semantic symbol vector ${\bf{X}}_n = [{\bf{x}}_{n,1}, {\bf{x}}_{n,2}, \dots, {\bf{x}}_{n,{k_n}L_n}]$, where ${\bf{X}}_n \in \mathbb{R}^{{k_n}L_n \times 2}$ and ${k_n}L_n$ is the length of the semantic symbol vector for a sentence at the $n$-th user. We notice that the length of ${\bf{X}}_n$ varies with $L_n$ to extract the semantic information of sentences with different lengths more effectively\cite{DeepSC}. In such a model, ${k_n}$ denotes the average number of semantic symbols used for each word at the $n$-th user, and each semantic symbol can be transmitted over transmission medium directly.
\vspace{-15pt}
\subsection{Transmission Model}
\vspace{-2pt}
Let $\mathcal{M}=\{1,2,\dots,m,\dots,M\}$ denote the set of available channels in the network, where $M$ is the number of channels and each channel is with bandwidth $W$. The channel assignment vector of the $n$-th user is denoted as ${\bm{\alpha}}_n = \{\alpha_{n,1},\alpha_{n,2},\dots,\alpha_{n,m},\dots,\alpha_{n,M}\}$, where $\alpha_{n,m}\in \{0,1\}$, $\alpha_{n,m}=1$ when the $m$-th channel is allocated to the $n$-th user, and $\alpha_{n,m}=0$, otherwise. Assuming that each channel can only be allocated to at most one user and each user can only occupy at most one channel, we have
\vspace{-3pt}
\begin{equation}
   \sum\limits_{n = 1}^N {{\alpha _{n,m}} \le 1,\ \forall m \in \mathcal{M}}; \sum\limits_{m=1}^M{{\alpha _{n,m}} \le 1,\ \forall n \in \mathcal{N}}.
   \vspace{-3pt}
\end{equation}

In addition, we consider that all channels consist of large-scale fading and small-scale Rayleigh fading. The signal-to-noise ratio (SNR) of the $n$-th user over the $m$-th channel is
\vspace{-3pt}
\begin{equation}
    {\gamma _{n,m}} = \frac{{p_n}{g_n}{|h_{n,m}|^2}}{W{N_0}},
    \vspace{-3pt}
\end{equation}
where $p_n$ is the transmit power of the $n$-th user, $g_n$ is the large-scale channel gain of the $n$-th user including path loss and shadowing, $h_{n,m}\sim{\mathcal{CN}(0,1)}$ is the Rayleigh fading coefficient for the $n$-th user transmitting over the $m$-th channel, and $N_0$ is the noise power spectral density.
\vspace{-12pt}
\subsection{DeepSC Receiver}
\vspace{-2pt}
At the BS, the signal from the $n$-th user can be denoted as 
${\bf{Y}}_n={\sqrt{g_n}}{h_{n,m}}{\bf{X}}_n+\bf{z}$ where ${\bf{z}}$ is Additive White Gaussian Noise (AWGN) and each element of ${\bf{z}}$ follows ${\mathcal{CN}(0,N_0)}$. The received signal will be decoded first by the channel decoder and thereby the semantic decoder to estimate sentence $\hat{\bf{s}}_n$.

In order to evaluate the performance of semantic communications for text transmission, we adopt the semantic similarity \cite{DeepSC} as the performance metric, 
\vspace{-3pt}
\begin{equation}
    {\xi} = \frac{{{\bf{B}}(s){\bf{B}}{{(\hat s)}^{\rm{T}}}}}{{\left\| {{\bf{B}}(s)} \right\|\left\| {{\bf{B}}(\hat s)} \right\|}},
    \vspace{-3pt}
\end{equation}
where $\bf{B}(\cdot)$ denotes Sentence-Bidirectional Encoder Representations from Transformers (BERT) model. It achieves great improvement over state-of-the-art sentence embedding methods. A pre-trained Sentence-BERT model \cite{reimers-2019-sentence-bert} is adopted. Compared with other semantic metrics, such as bilingual evaluation understudy (BLEU)\cite{bleu}, BERT-level similarity measures the distance of semantic information between two sentences more precisely. From (3), we have $0\le\xi\le1$ where $\xi=1$ means that two sentences has the highest similarity and $\xi=0$ indicates no similarity between them.

\vspace{-10pt}
\section{Semantic-Aware Resource Allocation}
\vspace{-3pt}
In this section, the S-SE is first defined as a new metric for semantic-aware networks. Then the semantic-aware resource allocation is formulated as a S-SE maximization problem in terms of channel assignment and the number of transmitted semantic symbols. Finally, the optimal solution of the optimization problem is obtained.
\vspace{-10pt}
\subsection{Semantic Spectral Efficiency}
\vspace{-3pt}
In conventional communications, spectral efficiency is 
measured in bits per second per Hertz (\textit{bits/s/Hz}), which can effectively measure the transmission rate of bit sequences but cannot be used to measure the transmission rate of semantic information. This is because the bit sequences are produced based on the statistical knowledge of the source and are irrelevant to the meaning of the source. Thus new performance metrics need to be investigated at the semantic level.

For the sake of clarity, we assume that semantic information can be measured by the \textit{semantic unit} (\textit{sut}), which represents the basic unit of semantic information\footnote{The \textit{semantic unit} here is just a concept and will not affect the resource optimization solution, the reason of which will be clarified in Section III-C.}. Based on this, two crucial semantic-based performance metrics can be defined:
\begin{itemize}
    \item {\textit{Semantic transmission rate (S-R)}} refers to the effectively transmitted semantic information per second and is measured in \textit{suts/s}.
    \item {\textit{Semantic spectral efficiency (S-SE)}} refers to the rate at which semantic information can be successfully transmitted over a unit of bandwidth, and is measured in \textit{suts/s/Hz}.
\end{itemize} 

Then the expressions of S-R and S-SE are derived respectively in the following. Denote $\mathcal{D} = \{({\bf{s}}_j=[w_{j,1}, w_{j,2}, \dots,w_{j,l}, \dots, w_{j,L_j}])\}_{j=1}^D$ with size $D$ as the text dataset, where ${\bf{s}}_j$ is the $j$-th sentence with length $L_j$ and $w_{j,l}$ is the $l$-th word. Let the amount of semantic information of ${\bf{s}}_j$ be $I_j$. With $p({\bf{s}}_j)$ representing the occurrence probability of ${\bf{s}}_j$, the expected amount of semantic information per sentence can be expressed as $I=\sum_{j=1}^D{{I_j}p({\bf{s}}_j)}$, which corresponds to an expected number of words per sentence as $L=\sum_{j=1}^D{{L_j}p({\bf{s}}_j)}$. Note that we focus on the long-term text transmission rather than the transmission of individual sentences, so the expected values $I$ and $L$, instead of the random values, should be taken to obtain the representations of S-R and S-SE. Hence, at the $n$-th user, there are ${k_n}L$ semantic symbols on average carrying the amount of semantic information of $I$, and the average amount of semantic information per semantic symbol is $I/({k_n}L)$. Moreover, since the symbol rate is equal to the channel bandwidth for passband transmission, the total semantic information transmitted over the channel with bandwidth $W$ is $WI/({k_n}L)$. Thus the S-R of the $n$-th user over the $m$-th channel can be expressed as
\vspace{-2pt}
\begin{equation}
    \Gamma_{n,m}=\frac{{W}I}{{{k_n}L}}{\xi_{\rm{n,m}}},
    \vspace{-2pt}
\end{equation}
where $\xi_{n,m}$ is the semantic similarity of the $n$-th user over the $m$-th channel. Note that $\xi_{n,m}$ relies on the neural network structure of DeepSC and channel conditions. It can be expressed as a function of $k_n$ and $\gamma_{n,m}$, i.e., $\xi_{n,m}=f(k_n,\gamma_{n,m})$. From (4), the corresponding S-SE can be expressed as
\vspace{-2pt}
\begin{equation}
    {\Phi_{n,m}}=\frac{\Gamma_{n,m}}{W}=\frac{I}{{{k_n}L}}{\xi_{\rm{n,m}}}.
    \vspace{-2pt}
\end{equation}
\vspace{-10pt}
\subsection{Problem Formulation}
\vspace{-2pt}
In this part, a semantic-aware resource allocation model is proposed to maximize the overall S-SE of all users. By denoting $\Phi$ as the overall S-SE of all users, we have
\vspace{-2pt}
\begin{equation}
    \Phi={{\sum\limits_{n = 1}^N {\sum\limits_{m = 1}^M {{\alpha _{n,m}}} }}}\frac{\xi_{n,m}I}{{k_n}L}.
    \vspace{-2pt}
\end{equation}

The channel assignment vector is considered as one of the optimization variables to fully exploit the performance advantage of DeepSC in the low SNR regime. Furthermore, we also optimize the average number of the transmitted semantic symbols for each word, $k_n$, to enable each symbol to carry more semantic information and thus achieve higher S-SE while ensuring the same transmission reliability.

According to the above analysis, the optimization problem can be formulated as
% \vspace{-2pt}
\begin{align}
    {\rm{\mathbf{(P0)}}}\quad\mathop{\max}\limits_{{{\bm{\alpha}}_n},{k_n}}\  &{\Phi}\label{YY}\\
    {\rm{ s.t.}}\ \ &{\rm{ C_1 }}:\  \alpha _{n,m} \!\in\! \{0,1\},\ \!\forall n \!\in\! \mathcal{N},\ \!\forall m \!\in\! \mathcal{M} \tag{\ref{YY}{a}}, \label{YYa}\\
    &{\rm{ C_2 }}:\  \sum\nolimits_{n = 1}^N {{\alpha _{n,m}} \le 1,\ \forall m \in \mathcal{M}}\tag{\ref{YY}{b}}, \label{YYb}\\
    &{\rm{    C_3   }}:\  \sum\nolimits_{m=1}^M{{\alpha _{n,m}} \le 1,\ \forall n \in \mathcal{N}} \tag{\ref{YY}{c}}, \label{YYc}\\
    &{\rm{     C_4  }}:\  {k_n} \in \{1,2,\dots,K\}\tag{\ref{YY}{d}}, \label{YYd}\\
    &{\rm{     C_5  }}:\   \xi_{n,m} \ge \xi_{\rm{th}}\tag{\ref{YY}{e}}, \label{YYe}\\
    &{\rm{C_6}}:\  \Phi_{n,m} \ge \Phi_{\rm{th}}\tag{\ref{YY}{f}} \label{YYf}
    \text{,}
    \vspace{-2pt}
\end{align}
where ${\rm{C_1}}$, ${\rm{C_2}}$, and ${\rm{C_3}}$ are channel assignment constraints, ${\rm{C_4}}$ specifies the permitted range of the average number of semantic symbols per word with $K$ representing the maximum value, ${\rm{C_5}}$ reflects the minimum required semantic similarity $\xi_{\rm{th}}$, and ${\rm{C_6}}$ restricts the minimum S-SE of users by $\Phi_{\rm{th}}$.  
\vspace{-10pt}
\subsection{The Optimal Solution}
\vspace{-2pt}
To solve ($\rm{\mathbf{P0}}$), two challenges should be addressed. One is how to deal with the term $I/L$ in the objective function, and the other is how to cope with $\xi_{n,m}$, which is closely related to $\Phi$, $\rm{C_5}$, and $\rm{C_6}$.

First, we note that the term $I/L$ depends on the type of source. According to the analysis in Section III-A, this term is a constant for a particular type of source, which will not affect the resource optimization. Consequently, we can omit this term when solving  ($\rm{\mathbf{P0}}$). Thus the optimization problem ($\rm{\mathbf{P0}}$) can be rewritten as 
\vspace{-2pt}
\begin{equation}
\begin{split}
    {\rm{\mathbf{(P1)}}}\quad\mathop{\max}\limits_{{{\bm{\alpha}}_n},{k_n}}\  &{\widetilde{\Phi}}={{\sum\limits_{n = 1}^N {\sum\limits_{m = 1}^M {{\alpha _{n,m}}} }}}\frac{\xi_{n,m}}{{k_n}}\\
    {\rm{ s.t.}}\ \ &{\rm{ C_1,C_2,C_3,C_4,C_5,C_6}},
\end{split}
\vspace{-2pt}
\end{equation}

Then, since $\xi_{n,m}$ is dependent of the specific semantic communication system and the physical channel conditions, we run the DeepSC model over AWGN channel to obtain the mapping between $\xi_{n,m}$ and $(k_n, \gamma_{n,m})$, as shown in Fig. 2. 

\begin{figure}
\vspace{-7pt}
  \centering
  \includegraphics[width=0.35\textwidth]{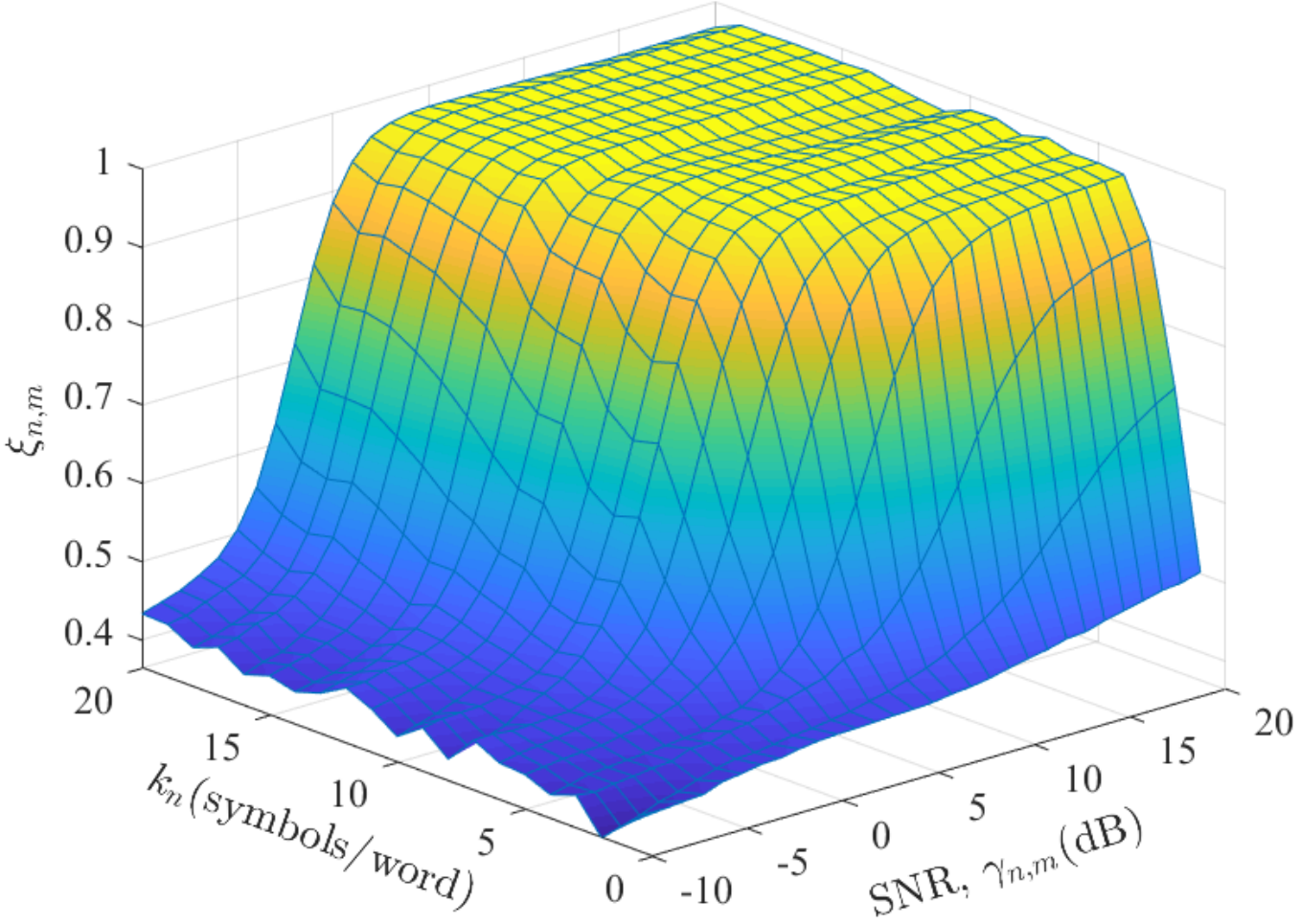}
\vspace{-8pt}
  \caption{The semantic similarity for DeepSC.}
  \vspace{-16pt}
\end{figure}
After addressing the two challenges, ($\rm{\mathbf{P0}}$) can be solved. Specifically, due to the orthogonality of different cellular links, ($\rm{\mathbf{P1}}$) can be decoupled into the following two equivalent independent optimization problems:
\vspace{-3pt}
\begin{equation}
\begin{split}
    {\rm{\mathbf{(P2)}}}\quad\mathop{\max}\limits_{{k_n}}\  &{\widetilde{\Phi}_{n,m}}\\
    {\rm{ s.t.}}\ \ &{\rm{ C_4,C_5,C_6}},
\end{split}
\vspace{-3pt}
\end{equation}
and
\vspace{-3pt}
\begin{equation}
\begin{split}
    {\rm{\mathbf{(P3)}}}\quad\mathop{\max}\limits_{{{\bm{\alpha}}_n}}\  &{\sum\limits_{n = 1}^N {\sum\limits_{m = 1}^M {{\alpha _{n,m}}} }\widetilde{\Phi}_{n,m}^{\rm{max}}}\\
    {\rm{ s.t.}}\ \ &{\rm{ C_1,C_2,C_3}}
    \text{,}
\end{split}
\vspace{-3pt}
\end{equation}
where ${\widetilde{\Phi}_{n,m}}={\xi_{n,m}}/{{k_n}}$ and $\widetilde{\Phi}_{n,m}^{\rm{max}}$ represents the maximum ${\widetilde{\Phi}_{n,m}}$ with respect to $k_n$. ($\rm{\mathbf{P2}}$) targets on obtaining ${\widetilde{\Phi}_{n,m}}$ for all users over all candidate channels. Since $\xi_{n,m}$ in $\rm{C_5}$ and $\rm{C_6}$ can only be obtained by the look-up table method, the exhausted searching method is adopted to solve ($\rm{\mathbf{P2}}$). Moreover, ($\rm{\mathbf{P3}}$) can be regarded as a maximum match problem of a bipartite graph. It can be solved by the Hungarian algorithm \cite{Hungarian}, where two vertex sets are $\mathcal{N}$ and $\mathcal{M}$ respectively, and ${\widetilde{\Phi}^{\rm{max}}_{n,m}}$ is regarded as the weight between the $n$-th user and the $m$-th channel.
\vspace{-12pt}
\section{Simulation Results and Comparison}
\vspace{-5pt}
In order to evaluate the performance of the proposed semantic-aware resource allocation scheme comprehensively, we conduct the following verifications in the simulation:
\begin{enumerate}
    \item Comparing the proposed resource allocation model against the conventional one to verify the proposed model in semantic-aware networks.
    \item Comparing the S-SE of semantic and conventional communication systems to show the superiority of semantic communications.
\end{enumerate}

Since the conventional systems are usually assessed in the bit domain, we first develop a transform method to convert the typical SE to the S-SE by taking the effect of source coding into consideration, making fair comparisons possible. On this basis, simulation results are presented and analysed.

\vspace{-9pt}
\subsection{The Transform Method for Fair Comparisons}
\vspace{-1pt}
In conventional communications, each letter in a word is mapped into bits through source encoder. From the semantic perspective, each bit can be loosely regarded as a semantic symbol although it may carry less semantic information than the semantic symbol of DeepSC. Similar to the definition in Section III-A, the equivalent S-R can be expressed as
\vspace{-5pt}
\begin{equation}
    \Gamma_{n,m}'=C_{n,m}\frac{I}{{\mu L}}{\xi_{n,m}},
    \vspace{-5pt}
\end{equation}
where $C_{n,m}$ is the transmission rate of the $n$-th user over the $m$-th channel, measured in \textit{bits/s}, and $\mu$ is defined as the transforming factor revealing the ability of the source coding scheme in compressing data, representing the average number of bits per word, measured in \textit{bits/word}. Specifically, if a word includes five letters on average and ASCII code is adopted to encode each letter, we will have $\mu=40$ bits/word. Moreover, when we assume no bit error in conventional communications, ${\xi_{n,m}}$ is equal to 1. By denoting $R_{n,m}=C_{n,m}/W$ as the SE, the equivalent S-SE can be given by 
\vspace{-5pt}
\begin{equation}
    \Phi_{n,m}'=R_{n,m}\frac{I}{{\mu L}}.
    \vspace{-5pt}
\end{equation}
Hence, the source coding process and bit transmission process are both considered to derive the S-SE of the conventional systems so that fair comparisons between different communication systems can be performed.
\vspace{-12pt}
\subsection{Benchmarks}
\vspace{-2pt}
Considering the proposed resource allocation scheme is for a specific semantic system, i.e., DeepSC, we compare it with the following three benchmarks, including an ideal system and two practical ones that have been widely deployed:
\begin{itemize}
    \item \textit{Ideal system}: Shannon limit can be achieved with no bit errors, i.e., $R_{n,m}={\rm{log_2}}(1+\gamma_{n,m})$.
    \item \textit{4G system}: According to the measured SNR, the BS obtains the channel quality indicator (CQI)\cite{4G-SNR-CQI}, based on which the achievable SE $R_{n,m}$ can be obtained according to Table 7.2.3-1 in 3GPP TS 36.213.
    \item \textit{5G system}: Similar to 4G, the BS gets CQI based on the measured SNR \cite{5GCQI}, and then obtains the achievable SE $R_{n,m}$ according to Table 5.2.2.1-2 in 3GPP TS 38.214.
\end{itemize}

Note that no scheme could achieve a higher bit transmission rate than the ideal system, but we focus on the S-SE to evaluate the performance in this paper. By adopting the developed transform method, the S-SE optimization problem of the above three benchmarks can be formulated as 
\vspace{-5pt}
\begin{align}
    {\rm{\mathbf{(P4)}}}\quad\mathop{\max}\limits_{{{\bm{\alpha}}_n}}\  &{\sum\limits_{n = 1}^N {\sum\limits_{m = 1}^M {{\alpha _{n,m}}} }\Phi_{n,m}'^\Delta}\label{RR}\\
    {\rm{ s.t.}}\ \ &{\rm{ C_1,C_2,C_3 }}, \nonumber\\
    &{\rm{C_7}}:\  \Phi_{n,m}'^\Delta \ge \Phi_{\rm{th}}\tag{\ref{RR}{a}} \label{RRa}
    \text{,}
    \vspace{-5pt}
\end{align}    
where $\Phi_{n,m}'^\Delta$ is the S-SE of the $n$-th user over the $m$-th channel in system $\Delta$, $\Delta \in \{\rm{Ideal,4G,5G}\}$. ${\rm{\mathbf{(P4)}}}$ can be solved by the method introduced in Section III-C. 
\begin{table}
\vspace{-3pt}
\renewcommand\arraystretch{1.2}
\centering
 \caption{Simulation parameters.}
 \vspace{-5pt}
  \begin{tabular}{|c|c|}
  \hline
  \textbf{Parameter} & \textbf{Value} \\
  \hline
 {Number of users}, $N$  & 5\\
 \hline
 {Number of channels, $M$}  & 5\\
 \hline
 {Channel bandwidth, $W$}  & 180 KHz\\
 \hline
 {Noise power spectral density, $N_{\rm{0}}$}  & -174 dBm/Hz \\
 \hline
 {Pathloss model} & 128.1+37.6lg[d(km)] dB \\
  \hline
 {Shadow effect factor} & 6 dB \\
 \hline
 {Transmit power, $p_{n}$}  & 10 dBm \\
 \hline
 {Maximum number of symbols per word, $K$} & 20 symbols/word\\
 \hline
 {Semantic similarity threshold, $\xi_{\rm{th}}$}  & 0.9 \\
 \hline
 {S-SE threshold, $\Phi_{\rm{th}}$} & 0.025($I/L$) suts/s/Hz\\
 \hline
 {Transforming factor, $\mu$} & 40 bits/word\\
 \hline
 \end{tabular}
 \vspace{-10pt}
\end{table}

\vspace{-10pt}
\subsection{Simulation Results}
\vspace{-5pt}
In our simulation, a circular network with radius $r=500$ m is considered where $N$ users are distributed uniformly. Unless specifically stated, the relevant parameters are listed in Table~I.  

\begin{figure}
  \centering
  \includegraphics[width=0.29\textwidth]{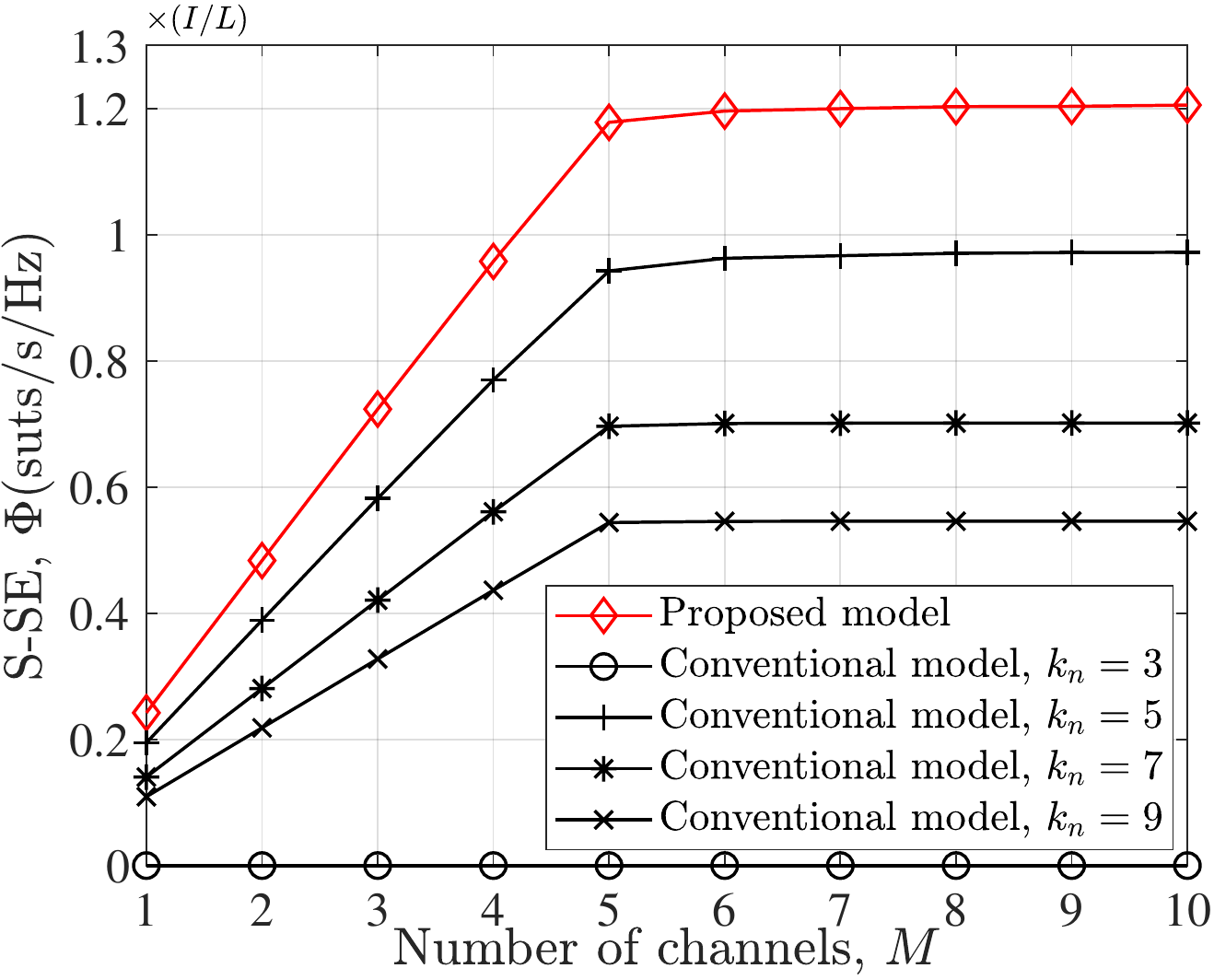}
\vspace{-10pt}
  \caption{The S-SE of the semantic-aware network with different models.}
\vspace{-18pt}
\end{figure}
We first examine the conventional resource allocation model in semantic-aware networks. In this simulation, the optimal channel assignment results of the conventional model in the ideal system is applied in the network, along with different values of $k_n$. Then the obtained S-SE is compared with that of the proposed model. 
As shown in Fig. 3, the S-SE of the conventional model is smaller than that of the proposed model regardless of the value of $k_n$, which implies that the conventional model is not suitable in semantic-aware networks. In addition, the S-SE of the conventional model with $k_n=3$ is equal to 0 because the semantic similarity is less than the threshold in this case.
\begin{figure*}[htbp]
	\centering
	\vspace{-3pt}
	\subfigure[The S-SE versus the number of channels.] {\includegraphics[width=.32\textwidth]{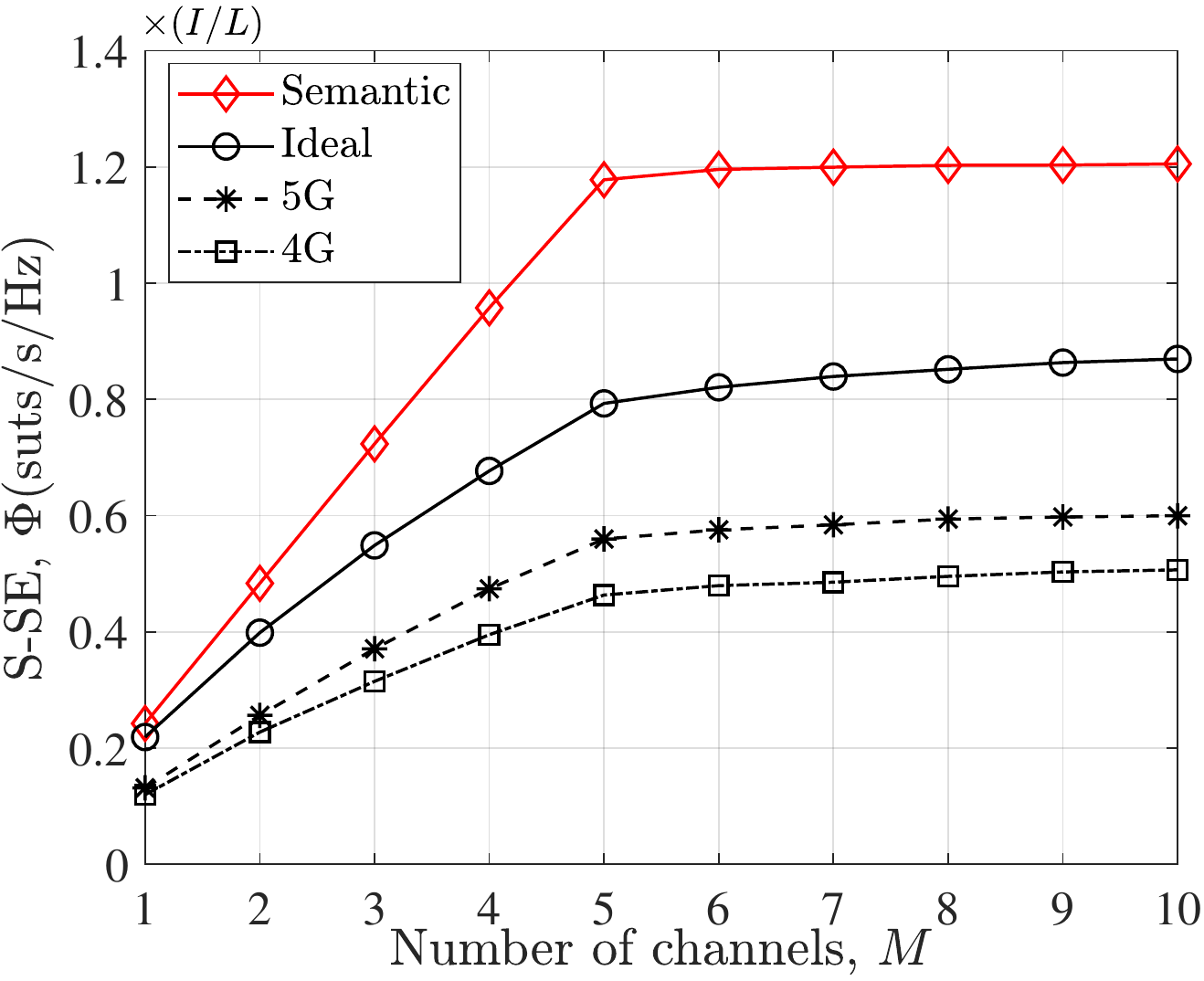}}
	\subfigure[The S-SE versus the transmit power.] {\includegraphics[width=.32\textwidth]{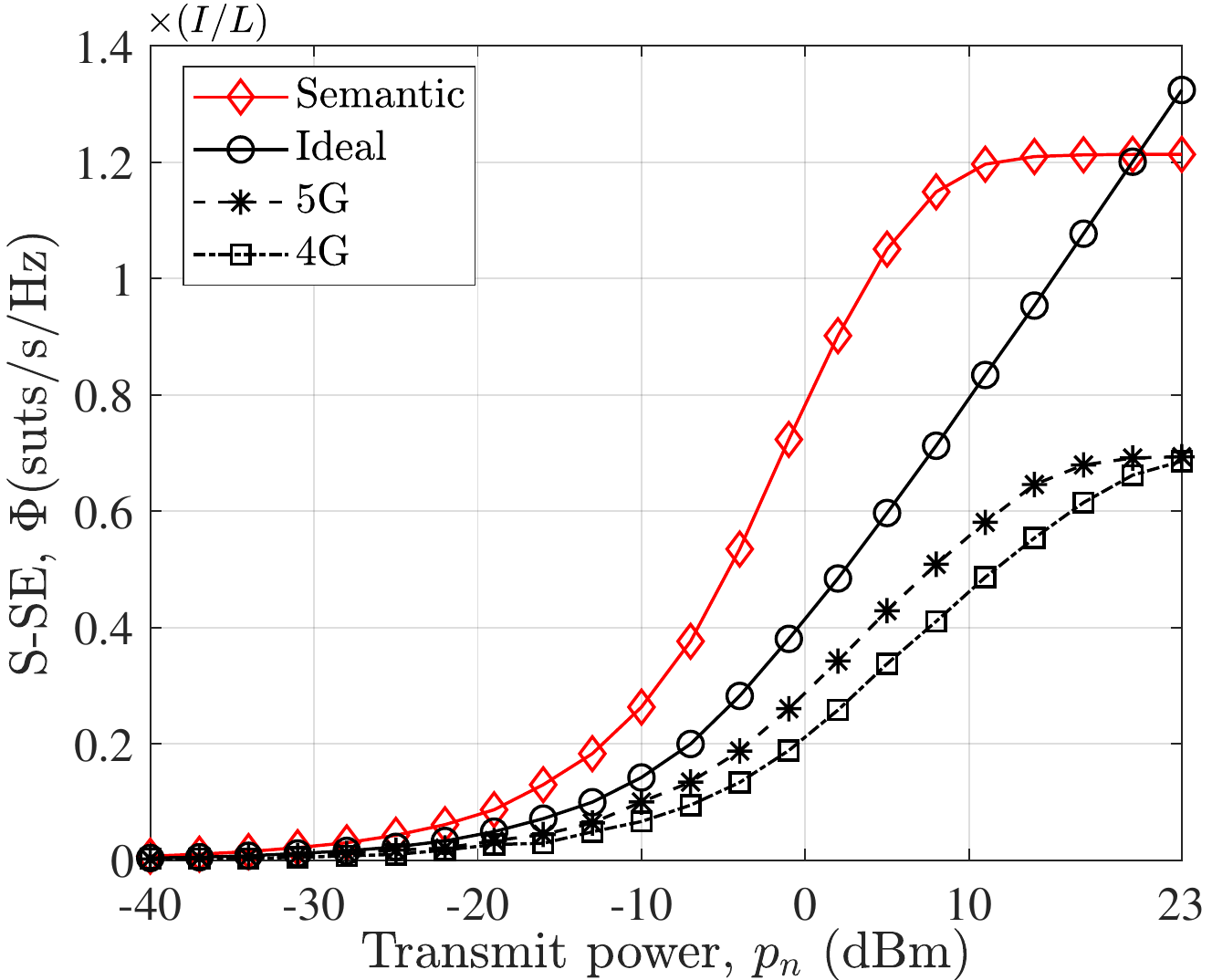}}
	\subfigure[The S-SE versus the transforming factor.] {\includegraphics[width=.32\textwidth]{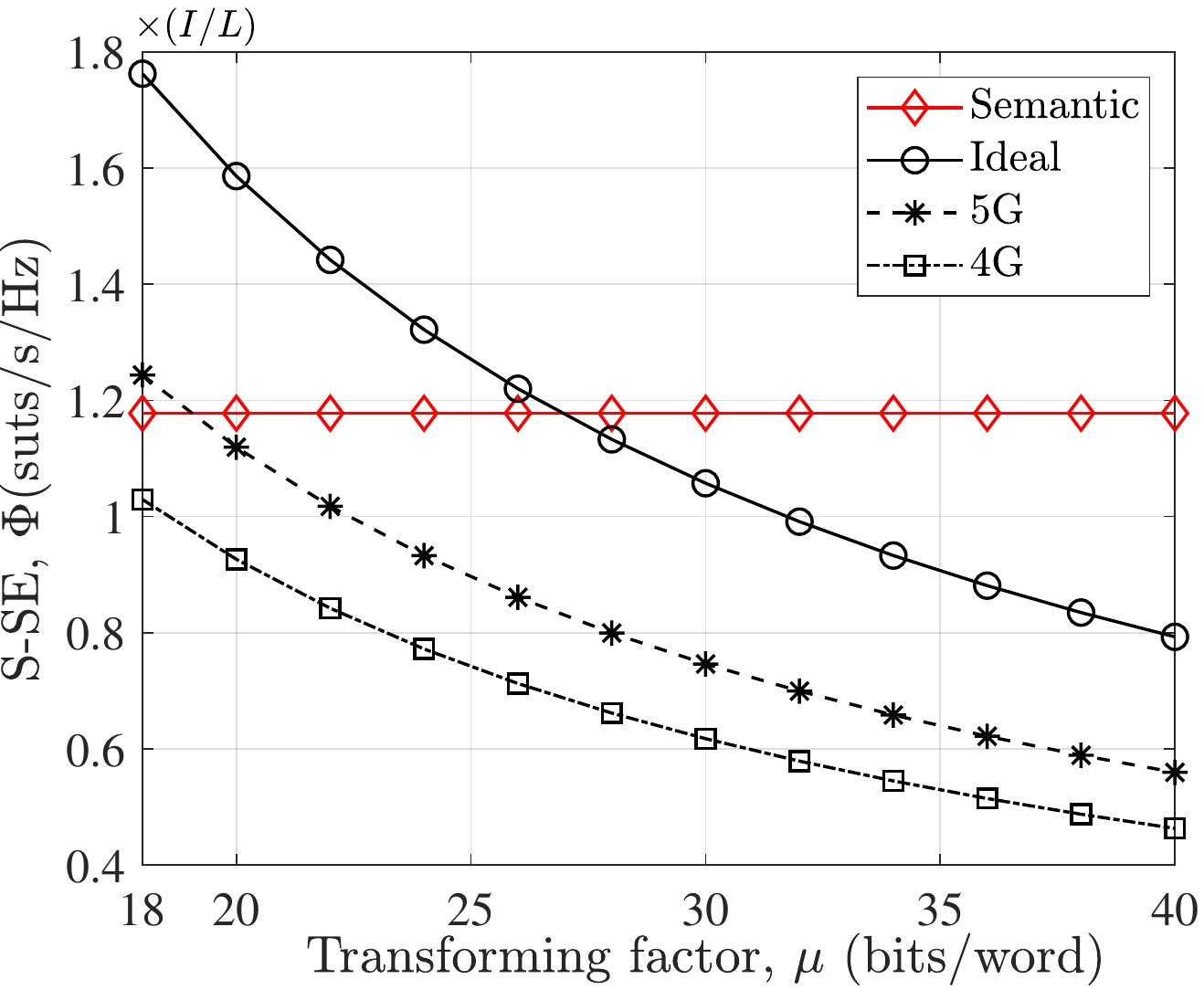}}
	\vspace{-8pt}
	\caption{The comparison of the semantic and conventional communication systems.}
\vspace{-16pt}
\end{figure*}

In the following, we compare the different communication systems with the corresponding resource allocation model. Fig. 4(a) shows the S-SE of different systems versus the number of channels. When $M$ is increased from 1 to 5, the S-SE of all systems increases rapidly because more users are served. Then when $M$ keeps on increasing from 5 to 10, the S-SE grows slowly instead of remaining stable because more channels are available and users can choose the channel with higher SNR. Moreover, the semantic communication system outperforms all conventional communication systems.

Fig. 4(b) illustrates the S-SE versus the transmit power. As $p_n$ increases, the S-SE of the ideal system increases rapidly while that of the semantic communication system, 4G system, and 5G system increase first and then tend to be a constant, implying that all practical systems have an upper bound with increasing SNR. Moreover, the semantic communication system shows a larger upper bound than 4G and 5G due to its stronger ability in compressing data.

Fig. 4(c) shows the S-SE versus the transforming factor. The performance of the semantic communication system remains stable since the transforming factor is irrelevant to it. For the conventional systems, the S-SE decreases with increasing $\mu$ because the S-SE is the ratio of the SE to $\mu$, and the maximum SE is a fixed value with different $\mu$. Additionally, the semantic communication system yields better performance than both 4G and 5G when $\mu$ is larger than 19 bits/word. Nevertheless, when $\mu$ is smaller than approximately 27 bits/word, i.e., a word can be encoded to less than 27 bits, the semantic communication system performs worse than the ideal system. This figure demonstrates that whether semantic communication systems outperforms conventional ones to a great extent depends on the source coding scheme adopted in conventional systems.

\vspace{-11pt}
\section{Conclusion}
\vspace{-1pt}
In this article, we have studied the SE issue in the semantic domain and explored the resource allocation for semantic communications. Specifically, S-R and S-SE have been defined first to make it possible to measure the communication efficiency of the semantic communication system based on the DeepSC model. Aiming at maximizing the overall S-SE of all users, the semantic-aware resource allocation has been formulated as an optimization problem and the optimal solution has been obtained. 
Extensive simulation has been conducted to evaluate the performance of the proposed scheme. An insightful conclusion is that, for text transmission, semantic communication systems achieve a higher S-SE than both 4G and 5G systems when a word is mapped to more than 19 bits on average through conventional source coding techniques. Further, if the required bits for encoding a word is increased to more than 27 bits with 10 dBm transmit power, semantic communication systems even outperforms the ideal system. In the future, how to design resource allocation method to satisfy the requirements of multiple intelligence tasks including single modal and multimodal tasks should be further investigated.
\vspace{-10pt}
\bibliographystyle{IEEEtran}
\bibliography{IEEEabrv}
\vspace{-3pt}
\end{sloppypar}
\end{document}